# Particle-induced transition in foams

Y. Khidas[1], B. Haffner[2] and O. Pitois[2]

[1] *Université Paris Est, Laboratoire Navier, UMR 8205 CNRS – École des Ponts ParisTech – IFSTTAR 5 bd Descartes, 77454 Marne-la-Vallée Cedex 2, France*

[2] *Université Paris Est, Laboratoire Navier, UMR 8205 CNRS – École des Ponts ParisTech – IFSTTAR cité Descartes, 2 allée Kepler, 77420 Champs-sur-Marne, France*

Abstract: The macroscopic behaviour of foams is deeply related to rearrangements occurring at the bubble scale, which dynamics depends on the mobility of the interstitial phase. In this paper, we resort to drainage experiments to quantify this mobility in particulate foams, where a particle suspension is confined between foam bubbles. Results show a strong dependence on each investigated parameter, i.e. bubble size, particle size and gas volume fraction for a given particle volume fraction. A combination of these parameters has been identified as the control parameter $\lambda$, which compares the particle size to the size of passage through constrictions within the foam pore space. $\lambda$ highlights a sharp transition: for $\lambda < 1$ particles are free to drain with the liquid, which involves the shear of the suspension in foam interstices, for $\lambda > 1$ particles are trapped and the mobility of the interstitial phase is strongly reduced.

Aqueous foams are dispersions of densely packed gas bubbles in liquid. Their structures are organized over a large range of length scales, which is the cause of the large variety of reported mechanical and dynamical behaviours [1]. They constitute a rich field of fundamental research [2] and they are often used as a model system for soft matter.

Foams are also used in a lot of industrial processes where gas is incorporated in large amounts to other components in order to produce new materials with improved functional properties, such as thermal performance for example. In this context, the optimization of such foamy materials requires a sound understanding of the general laws which control their behaviour. In spite of the significant progress realized in the field of foams, the results concerns almost entirely aqueous foams [1], whereas in industrial applications, complex fluids - such as suspensions - are mostly used as the continuous phase. Moreover, some very recent studies have highlighted the non-trivial behaviour of foams made with complex fluids. For example, the shear elasticity of foamy suspensions involves specific interactions between particles [3]. The drainage of foamy emulsions is activated by an applied macroscopic shear, according to a kinetic controlled by the magnitude of the shear rate [4]. From a more general point of view, the mechanics, the stability and the ageing of such systems involve bubble rearrangements, the so-called T1 events. During these rearrangements, the interstitial phase is redistributed within the bubble network according to a dynamic which has been studied exclusively in the context of aqueous foams [5-7]. For foams made with suspensions, the mobility of the interstitial phase accounts for the interplay between its bulk rheology and the interactions of particles with bubble interfaces. Unfortunately, this complex behaviour is still unexplained [8]. In this letter, we resort to drainage experiments to quantify the mobility of particle suspensions confined between the foam bubbles.

First we describe the procedure we have developed in order to produce controlled systems made of monodisperse particles, liquid and monodisperse bubbles. Using appropriate bubbling methods in a foaming solution (TTAB 10g/L, glycerol, water), a foam with bubble diameter $D_b$ is made in a vertical column. Liquid imbibition from the top of the column allows maintaining the liquid fraction at a constant value throughout the foam sample during the foam production. The foam is then pushed toward a T-junction where a suspension of polystyrene beads (diameter $d$ = 6-80 µm) is injected. The liquid phase is the same for the foam and for the suspension; its density was matched with that of polystyrene (1.05) by adjusting the proportion of glycerol (20% w/w) and its bulk viscosity is $\mu_0 \simeq 1.7$ mPa.s. The resulting gas and particle fractions, respectively $\phi$ and $\phi_p$, are set by the liquid fractions and the flow rates of injected foam and suspension. In the following, we will refer to the particle volume fraction in the interstitial suspension, i.e. $\varphi_p = \phi_p/(1-\phi)$. A systematic study of all parameters is performed for a given moderate concentration $\varphi_p = 0.16$. Besides, for a limited set of parameters we study the effect of $\varphi_p$ within the range 0 - 0.3. Our method has been found to produce homogeneous samples, characterized by well-distributed particles and bubbles, the size of the latter being preserved during the mixing step (Fig.1). The particle loaded foam is then continuously introduced in a rotating horizontal column used to compensate the effects of drainage during the preparation of the sample. Once the column is filled with the foamy suspension, it is turned to the vertical direction and the measurement of the free-drainage velocity starts. Note that with the present procedure, the starting point is a foam column with a uniform vertical gas fraction profile.

The mobility of the interstitial phase is determined via the drainage kinetics by following the height $h(t)$ corresponding to the volume of suspension drained off at the bottom of the column. Such a measurement is plotted in Fig. 2(a), showing a first stage characterized by a rapid linear increase of $h(t)$ for times $t < \tau$, followed by a slower evolution towards the equilibrium value $h_\infty$. Note that half of the liquid volume has drained off the foam as $t = \tau$ [1]. During this regime, the volume of liquid/suspension draining out of the foam flows through foam areas that are not yet reached by the drainage front, i.e. areas where the gas fraction remains equal to the initial value, $\phi$. We measure the velocity $V(\phi_s, \varphi_p, d, D_b)$ from the slope of this linear regime which accounts for drainage properties of the foam characterized by a constant gas fraction $\phi$, or equivalently by a constant volume fraction of suspension $\phi_s = 1 - \phi$. Note that although $\tau$ and $h_\infty$ vary significantly for particulate foams, it is shown in Fig. 2a that their drainage exhibits the same linear regime than the one described above for unloaded foams. In order to characterize the effect of particles on interstitial mobility, we normalize the measured drainage velocity by the one measured without particle, i.e. $V_0 \equiv V(\phi_s, \varphi_p = 0, d, D_b)$. Figs. 2(b), 2(c) and 2(d) show the reduced drainage velocity measured for several sets of parameters $(\phi_s, \varphi_p = 0.16, d, D_b)$, for which one parameter is changed as three others are fixed.

First of all, particles contained within the interstitial phase of the foam reduce systematically the drainage velocity with respect to unloaded foams. Whereas the velocity decreases significantly as a function of particle size, the opposite effect is measured for the bubble size. We also measure a strong influence of the volume fraction of suspension, which shows that the reduced particle size, i.e. $d/D_b$, is not an appropriate parameter to describe the drainage behavior of foamy suspensions. Thus we turn to another parameter, the so-called confinement parameter [9]:

$$\lambda = \frac{d}{d_c} = \frac{1 + 0.57\phi_s^{0.27}}{0.27\sqrt{\phi_s} + 3.17\phi_s^{2.75}} \frac{d}{D_b} \quad (1)$$

This geometrical parameter, which compares the particle size $d$ to the size $d_c$ of passage through constrictions within the foam pore space, has been determined from both experiments involving the trapping/release of a single particle in foams and numerical simulations of foam structures [9]. All the data presented in Fig. 2 are now plotted as a function of $\lambda$ in Fig. 3. The relative error is 14% for $\phi_s < 0.08$, and 6% for the other data. The data collapse satisfactory on a single curve meaning that $\lambda$ is the control parameter of this sharp transition from a regime $\lambda < 1$ where the reduced drainage velocity does not depend on the confinement to a regime $\lambda > 1$ where confinement leads to a severe drop of the mobility. In the following, we present the experimental and the theoretical arguments allowing to understand this non-trivial behaviour. As $\lambda$ has been identified as the control parameter, particle capture phenomena might play a crucial role in the interstitial mobility. Indeed, some samples release particles during drainage whereas others not. In order to quantify this effect, we measure the particle retention of each sample, i.e. the mass of particles caught by the foam after drainage divided by the mass of particles introduced into the sample. This procedure allows for the relative error to be 5-10% depending on the studied parameters. The results for the retention are presented in Fig. 4 as a function of the confinement parameter. The retention curve $R(\lambda)$ increases abruptly from 10% to 90% when $\lambda$ increases from 0.5 to 1.5. Note that during the free drainage, the gas fraction above the drainage front increases with time, resulting in the increase of the local confinement parameter in this upper part according to Eq. (1). This means that particles initially allowed to flow with the liquid can be trapped when the drainage front reaches them. This explains why $R(\lambda)$ can be non-null for $\lambda < 1$. As already explained, the volume of drained suspension ($v_\tau$) at $t = \tau$ equals the half of the final drained volume ($t \to \infty$), and the drainage conditions for $v_\tau$ are those set by the initial gas fraction. In other words, when $\lambda(\phi_s) \approx 1$, the 50% of particles released by the foam represent 100% of the particles contained in $v_\tau$ and the latter are released during the first regime of drainage. Since the measured drainage velocity corresponds to $v_\tau$, we define the proportion of trapped particles in $v_\tau$, i.e. $\xi_\tau(\lambda) = 2[R(\lambda) - 0.5]$ for $R(\lambda) > 0.5$ and $\xi_\tau(\lambda) = 0$ for $R(\lambda) < 0.5$ (inset Fig. 4). Our measurements reveal a progressive capture within the $\lambda$ range 0.9 - 1.7, whereas an ideal system would exhibit a step behaviour at $\lambda = 1$. This spread accounts for (i) the dispersion in the sizes of both channels and particles, (ii) the wall/bottom effects. Indeed, the wall Plateau borders are characterized by a $\lambda$ value 1.6 times larger than that corresponding to the bulk Plateau borders and their proportion is close on 10-15% [1].

In the understanding of the reported interstitial mobility transition, the key point is that the two drainage regimes correspond exactly to the $\lambda$ ranges where either $\xi_\tau \approx 0$ or $\xi_\tau \approx 1$.

Let first consider the case $\xi_\tau \approx 0$, which means that the particles are free to drain with the suspending liquid. In this regime, the reduced velocity appears to be constant. In the limit of vanishing $\lambda$, i.e. $\lambda \ll 1$, we expect the suspension to behave as a simple liquid with an effective reduced viscosity $\tilde{\mu}_{eff} = \mu_{eff}(\varphi_p)/\mu_0$. Here we refer to the semi-empirical relationship proposed by Krieger and Dougherty for the reduced effective viscosity of suspensions [14]. As the drainage velocity is inversely proportional to the liquid viscosity [1], for $\lambda \ll 1$ the reduced velocity writes $V/V_0 = 1/\tilde{\mu}_{eff}(\varphi_p) = (1 - \varphi_p/\varphi_p^*)^{2.5\varphi_p^*}$, where $\varphi_p^* \approx 0.6$ is the packing volume fraction of spherical particles. For $\varphi_p = 0.16$, one gets $V/V_0 \approx 0.63$, which is in very good agreement with all velocity values reported for $\lambda < 1$, even when $\lambda$ is close to unity. It should be noted that the volume of a foam node $v_n$ is large enough to be a representative volume of the suspension. For the rather wet foams considered here, most of the liquid/suspension is contained within the foam nodes [10] and $v_n$ can be estimated as follows: foam counts approximately 6 nodes per bubble [1] so that the node volume reads $(1 - \phi)/\phi \cdot \pi D_b^3/36 \approx 10^{-2} D_b^3$ for $\phi = 0.9$. Using Eq. (1), one gets $v_n \approx 30d^3$, which corresponds approximately to 60 sphere volumes. This means that although the geometrical confinement is extreme in the constrictions of the foam network for $\lambda \approx 1$, the concept of effective viscosity makes sense in foam nodes where the suspension is effectively sheared. Moreover, this effect is specific to foams due to the interfacial mobility which allows the particles to flow easily in constrictions [15].

Now we consider the regime $\xi_\tau \approx 1$, where the low value of the drainage velocity is due to the presence of captured particles. At the microscopic level we consider an effective foam node where the trapped particles are packed at the volume fraction $\varphi_p^*$. As already mentioned, the volume of the foam constrictions is neglected. During the drainage stage, the volume fraction of nodes filled with packed particles is $\Phi_n = \varphi_p/\varphi_p^*$. The pressure gradient over a loaded node is estimated by summing the pressure gradient over the portion of unloaded node [10] and the pressure gradient resulting from the permeation of the liquid flow through the porosity of the particle packing [11]: $\nabla P(\Phi_n) \approx (1 - \Phi_n)\mu_0 V_\ell/\tilde{k}_n r_{Pb}^2 + \Phi_n \mu_0 V_\ell/\tilde{C}_{CK} d^2$, where $V_\ell$ is the liquid velocity through the node, $\tilde{k}_n$ is the permeability coefficient of the node without particles, $\tilde{C}_{CK} \approx 10^{-3}$ [12] is the permeability coefficient for packed beads, and $r_{Pb}$ is the typical size of foam channels and it will be used here as the typical size of nodes. Note that $r_{Pb}$ is related to the constriction size: $d_c = \alpha r_{Pb}$, where $\alpha = 2(2/\sqrt{3} - 1)$ [9]. By averaging over all orientations for the nodes, the foam permeability $K$ can be expressed as a function of the parameters defined at the local scale [10]. As the drainage velocity is proportional to the foam permeability, we obtain the expression for the reduced drainage velocity: $V/V_0 = [(1-\Phi_n) + \Phi_n \tilde{k}_n/\tilde{C}_{CK}\alpha^2\lambda^2]^{-1}$. The fitting parameter $\tilde{k}_n$ is found to be 1/300, which is fully consistent with values reported in literature [13].

Thus, both retention regimes $\xi_\tau \approx 0$ and $\xi_\tau \approx 1$ are well understood in terms of drainage velocity. Now we concentrate on the transition between these two regimes, where the nodes are expected to be progressively filled with packed particles as $\lambda$ increases. A pragmatic approach consists in using the experimental retention curve for evaluating the volume fraction of nodes filled with packed particles within the range $0.9 < \lambda < 1.7$, i.e. $\Phi_n(\lambda) = \xi_\tau(\lambda)\varphi_p/\varphi_p^*$. We approximate $\xi_\tau(\lambda)$ by the simple form: $\xi_\tau(\lambda) = 5\lambda/4 - 9/8$, as presented in the inset of Fig. 4. Therefore, within this $\lambda$ range, the system consists in a volume fraction $\xi_\tau\varphi_p$ of trapped particles and a volume fraction $(1-\xi_\tau)\varphi_p$ of free particles. In the absence of the detailed description of the flow at the microscopic level, we assume that the transition can be described at the macroscopic scale by adding the contribution of each set of particles. Therefore, the effective viscosity of the flowing suspension becomes $\tilde{\mu}_{eff} = \left(1 - (1-\xi_\tau(\lambda))\varphi_p/\varphi_p^*\right)^{-2.5\varphi_p^*}$. Using the same approach as presented above we obtain:

$$V/V_0 = \left[(1-\Phi_n(\lambda))\tilde{\mu}_{eff}(\lambda) + \Phi_n(\lambda)\frac{\tilde{k}_n}{\tilde{C}_{CK}\alpha^2\lambda^2}\right]^{-1} \quad (2)$$

Eq. (2) is plotted in Fig. 3, where it is found to describe reasonably the transition observed in the experimental data. In order to check the robustness of the model we present in the inset the comparison of Eq. (2) with the whole set of experimental data. The best fitting value for $1/\tilde{k}_n$ is 300, but a reasonable agreement is observed as this parameter varies within the range 250-350. In contrast, the function $\xi_\tau(\lambda)$ has a more sensitive effect on the computed values. This shows that the retention function is crucial in the understanding of the drainage behaviour.

To conclude, we have highlighted a sharp transition in the drainage kinetics of particulate foams. This behaviour has been proved to be controlled by the confinement parameter $\lambda$. The significance of $\lambda$ has been emphasized by the measurement of particle retention during the drainage process, providing the basis for modelling. Extension of this work should consider more complex situations such as polydisperse systems (bubbles and/or particles). As the drainage velocity accounts for the mobility of the interstitial suspension, the reported results go well beyond the scope of drainage. Indeed, this mobility is involved in bubble rearrangements, the so-called T1 events, undergone by foams during flows [16], ripening [7] or coalescence events [17]. Therefore, this particle-induced mobility transition in foams is expected to be a major element in the understanding of the global behaviour of particles and bubbles mixed suspensions.

We thank A. Lemaître for stimulating discussions, D. Hautemayou and C. Mézière for technical support. We gratefully acknowledge financial support from Agence Nationale de la Recherche (Grant No. ANR-13-RMNP-0003-01) and French Space Agency (convention CNES/70980).

Figure 1

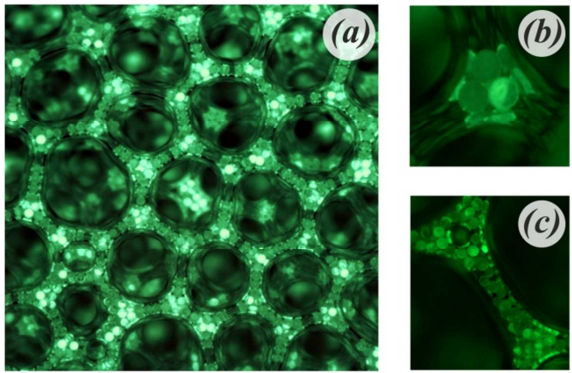

Fig. 1 (color online): Pictures showing foamy suspensions. (a) first layer of bubbles at the container wall. Fluorescent particles have been used in order to reveal the homogeneity of the sample over several bubble layers into the bulk. Bubble and particle sizes are respectively 660 µm and 80 µm ($\lambda = 1.7$). (b) detail from the left picture (in the bulk, not at the wall). (c) same situation except for the particle size (40 µm, $\lambda = 0.85$).

Figure 2

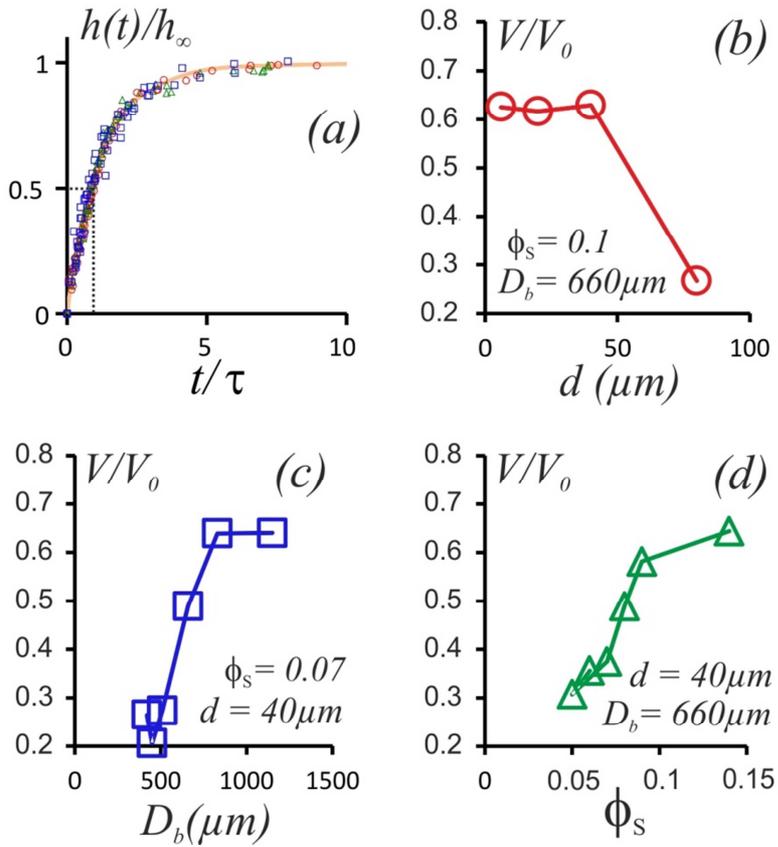

Fig. 2 (color online): Drainage velocity. (a) Temporal evolution of the reduced height of liquid/suspension drained out of the foam: unloaded foam (line), loaded foams (same symbols than in (b,c,d). Reduced drainage velocity as a function of particle size (b), bubble size (c) and volume fraction of suspension (d). For each sample $\varphi_p = 16\%$.

Figure 3

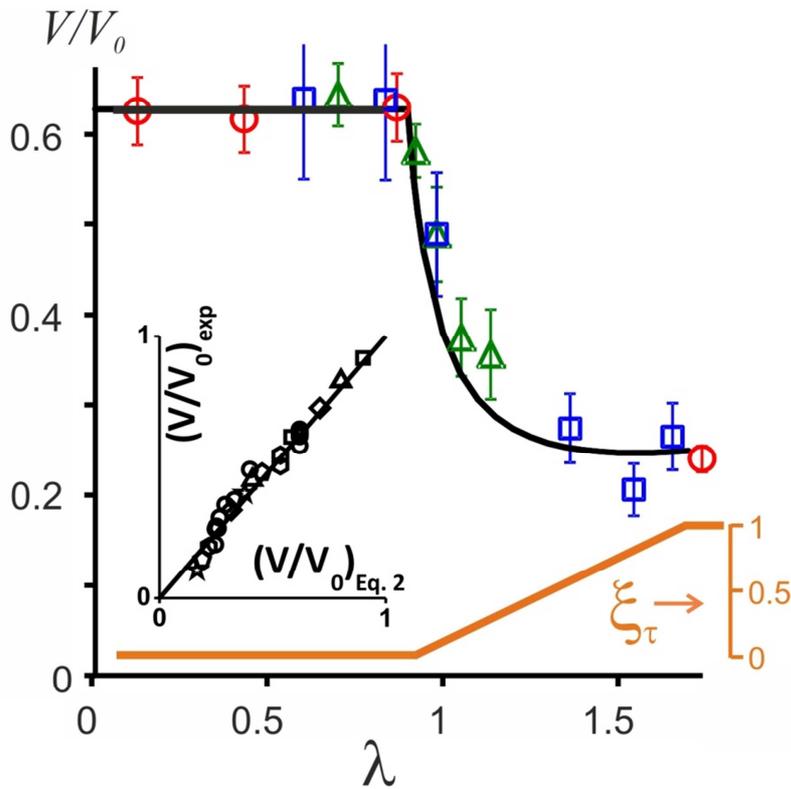

Fig. 3 (color online): Reduced drainage velocity as a function of the confinement parameter $\lambda$. The symbols refer to bubble sizes, particle sizes and volume fraction of suspensions presented in Fig. 2. The black line corresponds to Eq. 2. Bottom: particle retention. Inset: experiment vs Eq. 2 for 0.04 (□), 0.08 (△), 0.12 (◇), 0.16 (○), 0.2 (○), 0.25 (⬠) and 0.29 (★).

Figure 4

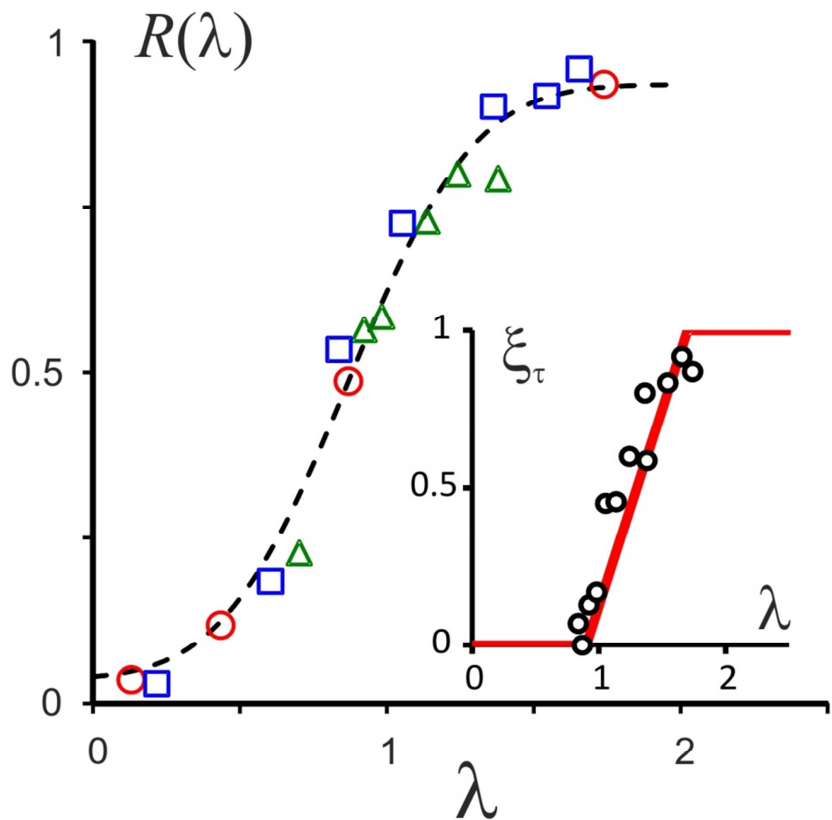

Fig. 4 (color online): Proportion of particles caught by foams after drainage as a function of the confinement parameter, for all investigated systems. Inset: Function $\xi_\tau$ defined as the proportion of trapped particles during the first drainage regime, i.e. for times $t < \tau$ (see fig. 2).